# Turnkey photonic flywheel in a Chimera cavity

Mingming Nie[1,*], Kunpeng Jia[2], Jan Bartos[1], Shining Zhu[2], Zhenda Xie[2], and Shu-Wei Huang[1,*]

[1]Department of Electrical, Computer and Energy Engineering, University of Colorado Boulder, Boulder, Colorado 80309, USA

[2]National Laboratory of Solid State Microstructures, School of Electronic Science and Engineering, College of Engineering and Applied Sciences, School of Physics, and Collaborative Innovation Center of Advanced Microstructures, Nanjing University, Nanjing 210093, China

[*]Corresponding author: mingming.nie@colorado.edu, shuwei.huang@colorado.edu

**Abstract:** Dissipative Kerr soliton (DKS) microcomb has emerged as an enabling technology that revolutionizes a wide range of applications in both basic science and technological innovation. Reliable turnkey operation with sub-optical-cycle and sub-femtosecond timing jitter is key to the success of many intriguing microcomb applications at the intersection of ultrafast optics and microwave electronics. Here we propose a novel approach to demonstrate the first turnkey Brillouin-DKS frequency comb. Our approach with a Chimera cavity offers essential benefits that are not attainable previously, including phase insensitivity, self-healing capability, deterministic selection of DKS state, and access to the ultralow noise comb state. The demonstrated turnkey Brillouin-DKS frequency comb achieves a fundamental comb linewidth of 100 mHz and DKS timing jitter of 1 femtosecond for averaging times up to 56 μs. The approach is universal and generalizable to various device platforms for user-friendly and field-deployable comb devices.

## Introduction

Dissipative Kerr soliton (DKS) frequency comb, generated by pumping an ultrahigh-quality-factor resonator, has been a ground-breaking technology with a remarkable breadth of demonstrated applications [1,2]. DKS frequency comb provides an easy access to large comb spacing in nonconventional spectral ranges, enabling high-capacity communication with high spectral efficiency [3,4], ultrafast optical ranging with massive parallelism [5], and high-speed spectroscopy in the molecular fingerprinting region [6].

In time domain, DKS timing jitter is the key property that determines its applicability as an optical flywheel [7–9], where the pristine temporal periodicity with sub-optical-cycle timing jitter can be utilized for demanding applications at the intersection of ultrafast optics and microwave electronics, including photonic analog-to-digital converters (ADCs) for next-generation radar and communication systems [10–12], ultrafast sub-nanometer-precision displacement measurement for real-time probing optomechanics, ultrasonic phenomena, and cell-generated forces [13,14], coherent waveform synthesizers for pushing the frontiers of femtosecond and attosecond science [15–17], and timing distribution links for synchronizing large-scale scientific facilities like X-ray free electron lasers and the extreme light infrastructure [18–22].

Sub-optical-cycle and sub-femtosecond timing jitter has long been theoretically predicted to be the DKS timing jitter's quantum limit [23]. However, due to the excessive technical noises, such quantum limit was not achieved until the two-step pumping scheme (see Extended Data Fig. 1) was recently invented to mitigate the pump-to-comb noise conversion and fundamentally lower the DKS timing jitter towards the quantum limit [8]. The two-step pumping scheme is a universal principle that utilizes Brillouin effect and enables the free-running photonic flywheel demonstration in various platforms including monolithic fiber Fabry-Pérot (FP) cavity [8,9], silica disk resonator [24] and silica wedge resonator [25]. From the quantum noise perspective, fiber cavity is a superior platform as its weaker mode confinement and lower Kerr nonlinearity preferably minimize the quantum limit towards sub-femtosecond DKS timing jitter in microsecond time scale [8,9,26]. The state-of-the-art Brillouin-DKS frequency comb demonstrated with the two-step pumping scheme achieves a fundamental comb linewidth of 400 mHz and DKS timing jitter of 1 femtosecond for averaging times up to 83 μs [9].

The remaining major obstacle to the widespread application of the Brillouin-DKS frequency comb is the lack of turnkey operation that mitigates the complex comb initiation dynamics and eliminates sophisticated feedback electronics (see Extended Data Fig. 1) to achieve compact footprint, low power consumption, and environmental ruggedness for the long-sought-after goal of user-friendly and field-deployable comb devices.

Here, we devise and demonstrate the first turnkey Brillouin-DKS frequency comb by combining the two-step pumping scheme and active laser gain that realizes decoupling between the pump generation and the comb generation. We build a Chimera cavity that consists of a comb-generating passive FP microresonator nested in a pump-generating active ring cavity (Fig. 1a). The Chimera cavity configuration utilizing the pump-comb decoupling characteristic distinguishes our approach from the nonlinear self-injection locking (SIL) method implemented to put the conventional DKS frequency comb into the turnkey operation regime [27–29]. Unlike nonlinear SIL, our approach does not suffer from the vulnerability to feedback phase fluctuation and enables the deterministic selection of

DKS soliton numbers. In addition, we demonstrate the self-healing capability of returning to the original comb state from instantaneous perturbations. More importantly, our approach allows the access to the ultralow noise comb state. The turnkey Brillouin-DKS frequency comb achieves a fundamental comb linewidth of 100 mHz and DKS timing jitter of 1 femtosecond for averaging times up to 56 μs, rendering it suitable for photonic ADCs, ultrafast optical ranging, coherent waveform synthesizers, and timing distribution links.

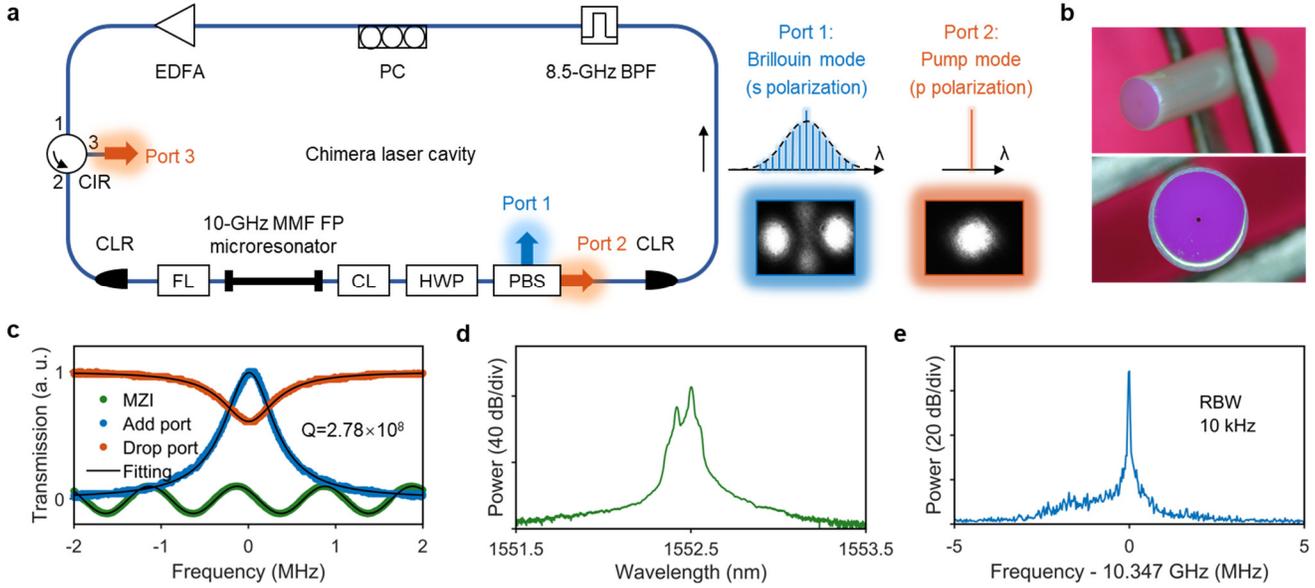

**Fig. 1. Brillouin-DKS generation in a Chimera cavity**. (a) Left: Schematic of the Chimera cavity that consists of a ~10-GHz multimode fiber (MMF) FP microresonator nested in a single-mode fiber (SMF) ring laser cavity. EDFA: Erbium-doped fiber amplifier, CIR: circulator, CLR: collimator, FL: focusing lens, CL: collimating lens, HWP: half-wave plate, PBS: polarization beam splitter, BPF: band-pass filter, PC: polarization controller. Right: spectra and beam profiles from PBS Port 1 and Port2. (b) Photographs of the MMF FP microresonator. (c) Frequency-calibrated transmission spectra of the pump mode, showing the microresonator linewidth of 695 kHz. (d) Optical spectrum output from PBS Port 1, including the Brillouin mode and leaked weak-intensity pump mode. (e) Radio frequency (RF) beat note of the stimulated Brillouin scattering (SBS) frequency shift at 10.347 GHz.

## Results

**A Chimera cavity.** Figure 1a shows the schematic of the Chimera cavity that consists of a comb-generating passive FP microresonator nested in a pump-generating active ring cavity. The FP microresonator is made of graded-index multimode fiber (GRIN-MMF) (Fig. 1b), and its linewidth, quality factor, and free spectral range (FSR) are 695 kHz, $2.78 \times 10^8$, and 10.087 GHz respectively (Fig. 1c, also see Methods). The SMF active ring cavity has a FSR of 26.8 MHz, much larger than the microresonator linewidth, and the embedded BPF has a bandwidth of 8.5 GHz, narrower than the microresonator FSR (see Methods). This arrangement ensures single frequency laser oscillation in the active ring cavity and such single-frequency laser pump is coupled into the fundamental mode of the FP microresonator for intermodal excitation of cross-polarized stimulated Brillouin lasing (SBL, Figs. 1d and 1e) that in turn generates the DKS microcomb through the two-step pumping scheme [8,9]. The single-frequency laser pump and the cross-polarized SBL can be straightforwardly separated by a PBS as shown in Fig. 1a to achieve the required decoupling between the pump generation and the comb generation. Of note, the Chimera cavity shares some similarity with the self-emergent laser cavity-soliton (LCS) [30] but the working principles are distinctly different. Self-emergent LCS critically depends on the parameters of the outer laser cavity which is more susceptible to environmental disturbance. Ultralow comb noise and long-term stability have not been demonstrated with the self-emergent LCS approach.

**Principle of turnkey Brillouin-DKS in a Chimera cavity.** In the two-step pumping scheme, as the pump frequency is swept from the blue-detuned side to near the resonance peak, SBL is first excited and then grows to generate Brillouin-DKS (see Extended Data Fig. 1). If the offset frequency between pump and Brillouin mode resonances is set to be slightly larger than the SBS frequency shift, blue-detuned pump and red-detuned SBL can simultaneously exist in the microresonator. Their opposite thermal nonlinearities can compensate each out and render such Brillouin-DKS thermally stable and manually accessible with expanded existence range. In the Chimera cavity, the pump-generating active ring cavity keeps the pump detuning around the resonance peak as the lasing will self-organize itself to the minimum loss and the maximum gain [31]. When the microresonator temperature (MRT) is set correctly such that the offset frequency is slightly larger than the SBS frequency shift, the co-existing blue-detuned pump and red-detuned SBL in the microresonator manifests itself into the thermally stable DKS attractor. MRT is thus an effective control parameter that deterministically select the DKS soliton number (see Fig. 3 and Supplementary Information).

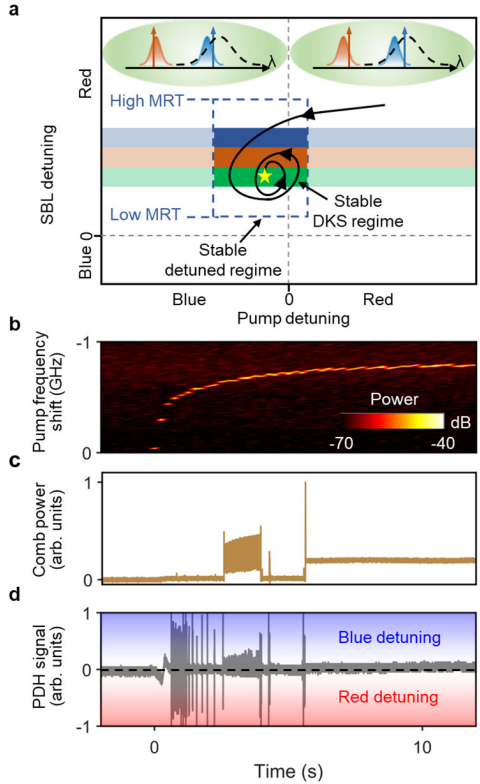

**Fig. 2. Principle of turnkey Brillouin-DKS generation.** (a) The DKS attractor is defined by the intersection of thermally stable regime (dashed box) and SBL-detuning controlled DKS regime where the soliton number is color coded. Of note, the microresonator temperature (MRT) changes the offset frequency between pump and Brillouin mode resonances and consequently MRT is an effective control parameter that determines the DKS attractor. Left inset: at the final stable DKS state (indicated by the yellow star), pump is blue detuned while SBL is red detuned. The black dashed line represents the Brillouin gain spectrum. Right inset: at an intermediate unstable state, both pump and SBL are red detuned. Evolution of (b) pump frequency shift, (c) comb power and (d) pump PDH signal during the turn-on process. PDH signal>0: blue detuning, PDH signal<0: red detuning, PDH signal=0: zero detuning (resonance peak).

While the DKS attractor state is mainly defined by the microresonator, the turnkey dynamics closely follows the optical pathlength change of the active ring cavity that is caused by refractive index increase and thermal expansion resulting from the EDFA pump absorption [32–34]. Figure 2b plots the pump frequency red shift during the laser turn-on process, which serves as a spontaneous scanning of pump detuning to drive the system evolution into the DKS attractor state. Figures 2c and 2d show the synchronously measured evolution dynamics of comb power and pump Pound–Drever–Hall (PDH) signal (see Methods), respectively. After several oscillations, stable DKS eventually forms when the pump laser is clamped at the slightly blue-detuned side of the microresonator resonance (Fig. 2d).

**Phase-independent turnkey Brillouin-DKS generation with deterministic soliton numbers.** To demonstrate repeatable turnkey operation, the 980-nm laser of the EDFA is modulated by a chopper with a square-wave profile to mimic the turn-on process. As shown in Fig. 3a, soliton microcomb operation is reliably achieved, as confirmed by synchronously monitoring the comb power and the clean RF beat note of the comb repetition rate. When the fiber cavity length is either fine-tuned with a resolution of 0.2 μm in a range of 10 μm or coarsely tuned with a resolution of 1 mm in a range of 20 mm, near 100% turnkey success probability can be achieved (Fig. 3b), indicating phase-independent and environment-insensitive turnkey operation, which is user-friendly and in sharp contrast to the nonlinear SIL method [27–29].

Figure 3c shows the optical spectra of a single-soliton state and perfect soliton crystal states with two and three solitons, corresponding to repetition rate of 10.09 GHz, 20.18 GHz and 30.26 GHz, respectively. The RF beat notes of comb repetition rate with high contrast and single tone (right insets of Fig. 3c) indicate the stable mode-locking status. Of note, different soliton states are achieved via slightly changing the microresonator temperature by ~0.2 K thus changing the final stable SBL detuning when the system reaches thermal equilibrium. In addition, as shown in the left insets of Fig. 3c, the system can robustly evolve into the same soliton states with probability of >90%, indicating a deterministic turnkey process. The deterministic perfect soliton crystal sates are believed to be caused by either the pump-SBL pair induced cross-phase modulation [35] or two co-lasing lasers due to the insufficient filtering effect of the BPF during the turnkey process. Of note, our deterministic turnkey process is independent of 980-nm laser power of the EDFA in an offset range of ±100 mW, where the reported single LCS attractor can be destroyed [30].

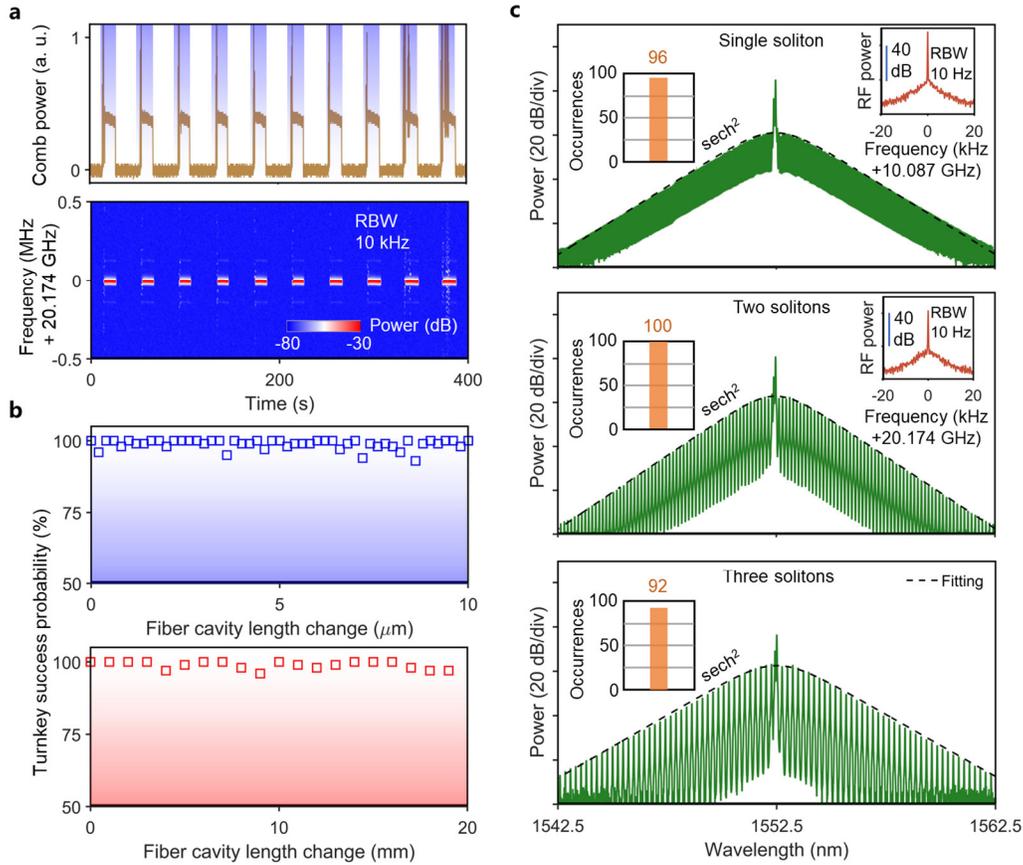

**Fig. 3. Phase-independent turnkey Brillouin-DKS generation with deterministic soliton numbers.** (a) Ten consecutive switching tests. The top panel shows the measured evolution of comb power in arbitrary units (a. u.), while the bottom panel shows the measured electrical spectrum of the soliton repetition rate during the switching process. The EDFA is switched on periodically, as indicated by the shaded regions. (b) Turnkey success probability by changing fiber cavity length with fine tuning (top) and coarse tuning (bottom). Each data point is acquired from 100 switch-on attempts. (c) Optical spectra of the turnkey SBL solitons. The black dashed lines show the fitted soliton spectral envelope. Left insets are the occurrences of single soliton (top), two solitons (middle) and three solitons (bottom) among 100 times soliton generation. Right insets are the clean and high-contrast RF beat notes of comb repetition rate.

**Strong immunity to perturbation.** The generated Brillouin-DKS in our Chimera cavity is strongly immune to the environmental perturbation, including the fiber cavity length change, the vibration and temperature change. Figure 4a shows the soliton self-healing from instantaneous fiber cavity length change of 1 μm for the two-soliton state in Fig. 3c. Both the comb power and soliton repetition rate recover to the original state after the perturbation. Besides, the Brillouin-DKS can keep stable within a cavity length change of ±4 μm at a low evolving speed at 0.1 Hz, corresponding to soliton repetition rate change of ±4.5 kHz (Fig. 4b). At the same time, the pump laser frequency shift is ±60 MHz while the pump detuning change is estimated to be ±100 kHz. Extended Data Fig. 2 also shows soliton self-healing from kicking the optical table and heating the fiber Bragg grating (FBG) based BPF by 1 K.

**Excellent noise performance.** Figure 5a compares the single sideband (SSB) frequency noise spectra of the lasing pump mode, SBL and SBL comb lines, measured with a low-noise-floor optical frequency discriminator (see Methods). Thanks to the ultrahigh Q of the MMF FP microresonator, all the lasing pump mode, SBL and SBL comb yield a fundamental linewidth of ~100 mHz, representing the record-breaking narrowest DKS comb linewidth and approaching that of the state-of-the-art on-chip SBLs [36–38], but now at free running without any active electronics. In addition, the relative intensity noises (RINs) of the lasing pump mode, SBL and SBL combs are all below -120 dB/Hz above 1 kHz offset frequencies (Fig. 5b). Of note, 180-mW high power from Port 3 (Fig. 1a) is achieved for the single-frequency pump due to the non-critical coupling (Fig. 1c) and ~70% external coupling efficiency. The SBL combs as well as the high-power lasing pump with low frequency noise and low RIN can benefit applications including coherent optical communications and optical atomic clocks.

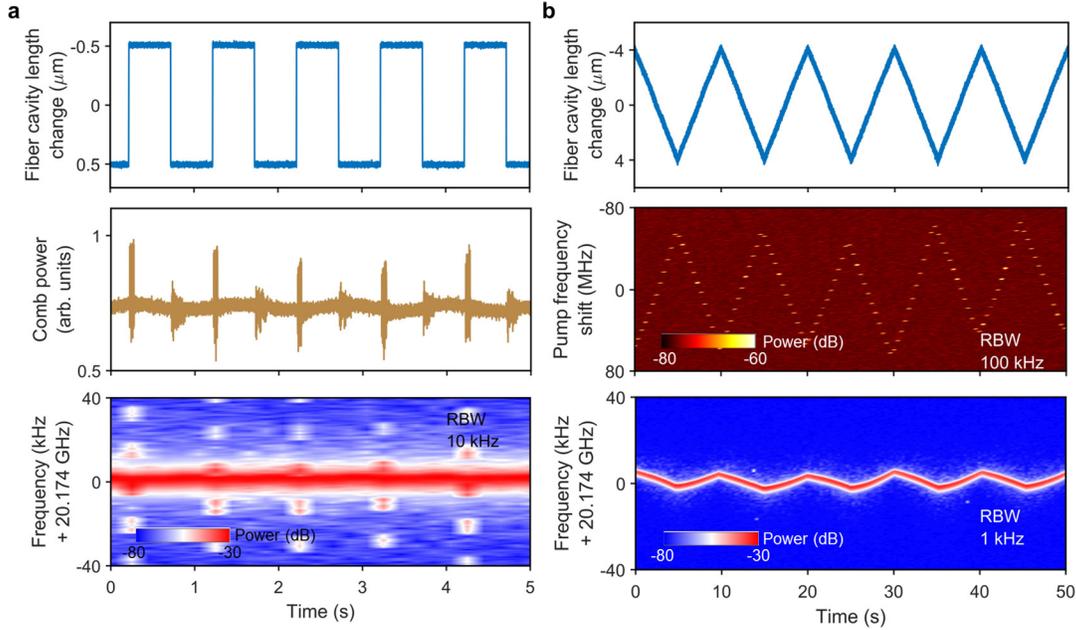

**Fig. 4. Strong immunity to perturbation.** (a) Soliton self-healing from instantaneous perturbation to the fiber cavity length. (b) Soliton immunity to the slow modulation of the fiber cavity length.

Owing to the narrow linewidth for both the lasing pump and the SBL, the achieved SBS frequency shift at 10.347 GHz is a good frequency synthesizer with low SSB phase noise of -107 dBc/Hz at 100 kHz (Fig. 5c), whose performance is comparable to the one utilizing cascaded Brillouin process [38,39] but now self-starts without pumped by an external-cavity diode laser (ECDL). In Fig. 5c, we also plot the SSB phase noise of the SBL soliton repetition rate, measured with the all-fiber reference-free Michelson interferometer (ARMI) setup providing attosecond timing jitter resolution [8,9,40,41] beyond the capability of direct photodetection methods [42–44] (see Methods). The measured SSB phase noises at 10 kHz, 100 kHz, and 1 MHz offset frequencies are -128 dBc/Hz, -147 dBc/Hz, and -166 dBc/Hz, respectively. The timing jitter integrated from 18 kHz to 1 MHz is 1 fs (see Extended Data Fig. 3), which is less than one-fifth of single optical cycle at the SBL soliton center wavelength, reaching the photonic flywheel level. Compared with the soliton phase noise using nonlinear SIL method [29,45,46] and the phase noise of the SBS frequency shift (blue line in Fig. 5c), our Brillouin-DKS phase noise not only represents improvement of -10 dBc/Hz per decade, following $1/f^2$ trend with the offset frequencies, but also reaches lower noise of -166 dBc/Hz at 1 MHz offset frequency. The ultralow soliton phase noise can benefit critical applications including photonic ADCs, ultrafast optical ranging, coherent waveform synthesizers and timing distribution links.

The long-term stability of the SBL soliton microcombs is also examined to be excellent in the laboratory environment (Fig. 5d). During an operating period of 2 hours, the standard deviation of comb power and soliton repetition rate for the two-soliton state are 0.25% and 1.38 kHz with a resolution bandwidth of 1 kHz. The pump frequency shift and the pump detuning shift are also measured to be small as 200 MHz and 70 kHz, respectively, as shown in Extended Data Fig. 4.

**Discussion**

Our Chimera cavity, where pump generation in the active cavity and DKS generation in the microresonator are decoupled, is a universal topology for turnkey DKS generation and has the potential for fully on-chip integration. Silicon nitride Vernier microring filter has been proven effective as a narrow on-chip BPF [47]. Heterogeneously integrated semiconductor optical amplifier [28] and Erbium-doped silicon nitride waveguide amplifier [48] have both been recently demonstrated as viable on-chip gain media. SBL generation has been achieved in a weakly-confined silicon nitride microresonator [38] and a recent study further shows the potential of a tightly-confined silicon nitride waveguide [49]. Moreover, SBS is not the only intracavity effect that can be used to implement the two-step pumping scheme. Avoided mode crossing (AMX) [50–53] can also be utilized and it relaxes the need to match the microresonator FSR with the SBS frequency shift, rendering AMX more flexible and user-friendly for the on-chip turnkey DKS microcomb implementation.

In conclusion, we demonstrate a novel and universal approach to mitigate previous challenges and achieve reliable turnkey DKS frequency comb generation. Phase insensitivity, self-healing capability, deterministic selection of DKS state, and access to the ultralow noise comb state are all successfully accomplished. The turnkey Brillouin-DKS frequency comb achieves a fundamental comb linewidth of 100 mHz and DKS timing jitter of 1 femtosecond for averaging times up to 56 μs, rendering it suitable for applications at the intersection of ultrafast optics and microwave electronics such as photonic ADCs, ultrafast optical ranging, coherent waveform synthesizers, and timing distribution links. The approach can be generalized to photonic integrated circuit platforms for mass production of user-friendly and field-deployable comb devices.

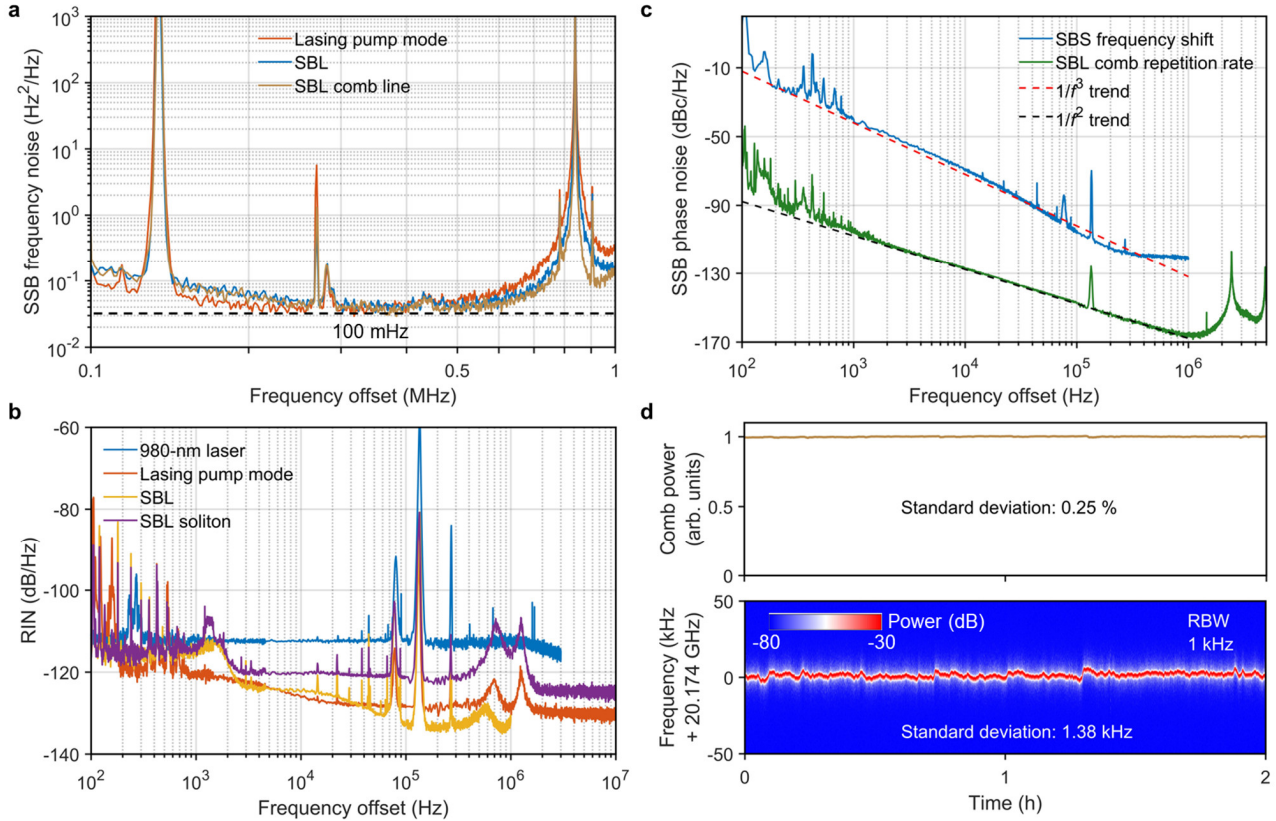

**Fig. 5. Excellent noise performance.** (a) SSB frequency noise spectra of the lasing pump mode, SBL and SBL comb line. Fundamental linewidths are calculated from the white noise floor of the measured SSB frequency noise spectra. The same fundamental linewidths of both pump laser and SBL and weak linewidth narrowing factor from SBS process are attributed to the limit of laser RIN (see Supplementary Information). (b) RIN of 980-nm laser, pump mode, SBL and SBL soliton. (c) SSB phase noise spectra of the SBS frequency shift (10.347 GHz) and repetition rate of single SBL soliton (10.087 GHz). The coherent artefacts at 2.5 MHz and its harmonics (green line) result from the 82-m delay fiber used in the ARMI setup, while the peak at 135 kHz comes from the EDFA pump RIN. (d) Long-term stability of the comb power and comb repetition rate over 2 hours in the laboratory environment with temperature variation of ±1 K.

## Methods
### MMF FP microresonator

Our high-Q MMF FP microresonator is fabricated through three steps: (i) commercial MMF (GIF50E, Thorlabs) is carefully cleaved and encapsuled in a ceramic fiber ferrule; (ii) both fiber ends are mechanically polished to sub-wavelength smoothness; (iii) both fiber ends are coated with optical dielectric Bragg mirror with reflectivity over 99.9% from 1530 to 1570 nm. The large mode area of the MMF leads to low diffraction losses in the thick dielectric Bragg mirror coatings at both ends and results in ultra-high Qs for both the internal pump mode and Brillouin mode [9].

The 10-mm long FP microresonator, corresponding to a OPL of ~29 mm, results in a cold-cavity free spectral range of 10.087 GHz, which is ready for microwave photonics applications. The group velocity dispersions of all the supported modes are simulated to be anomalous of ~-28 $fs^2$/mm [9].

The ultrahigh-Q microresonator acts as two roles here: (i) a high-finesse etalon filter in the fiber laser cavity responsible for single-frequency pump mode with low frequency noise; (ii) the device generating SBL and corresponding DKS.

**Details of experimental setup.** The fiber cavity includes 2-m gain giber, 5-m passive fiber and 1-m free space length, resulting in OPL of 11.2 m and a FSR of 26.8 MHz. The EDFA is home-made, consisting of a 2-m highly Erbium-doped fiber (SM-ESF-7/125, Nufern). The narrow BPF is consisted of a circulator and a temperature-controlled FBG centered at 1552.436 nm with 3-dB bandwidth of 0.068 nm (8.5 GHz). In order to achieve low coupling loss, free-space components are introduced to couple the light into and out from the MMF FP microresonator with one-end coupling efficiency of ~70%. We believe all-fiber integration is feasible due to the compatibility between the MMF FP microresonator and other fiber components. The EDFA part and narrowband filter part with 3-m total bare fiber is shielded but not temperature controlled while the other components are exposed to the lab environment including the 4-m passive fiber protected with 0.9-mm jacket. The optical circulator ensures the unidirectional lasing of the fiber cavity and couples out the high-power single-frequency pump laser. The polarization controller is used to maximize the coupling efficiency into the microresonator.

The Q of the Brillouin mode is not measured due to the difficulty of finding the high-order MMF mode and the low coupling efficiency between the single-mode fiber and MMF. However, we can still estimate its Q to be large than $1\times10^8$ according to our previous experiment data with the same kind of large-mode-area FP microresonators [9]. In addition, the total input pump mode power before coupling into the microresonator is as low as ~250 mW at 980-nm pump power of 1.5 W for the EDFA, also confirming the ultrahigh Q of the Brillouin mode. The MMF FP microresonator is temperature controlled with a resolution of 10 mK (see Supplementary Information for detials).

A phase modulator is inserted in the fiber cavity for all the experiments to monitor the detuning of the lasing pump mode via the powerful PDH technique. The phase modulator is inserted before the BPF (Fig. 1c). The modulation frequency is 1 MHz and the PDH signal is demodulated by the single-frequency pump laser output from the circulator. The low-pass filter used in the PDH signal demodulation is 100 kHz. The modulation voltage of the phase modulator is chosen to be low without perturbing the SBL soliton generation and turnkey operation. Thermal effect and Kerr effect are considered in the PDH signal measurement.

The pump laser frequency shift is monitored by the beat note between the pump laser and a tunable ECDL. The beat note is measured by an electrical spectrum analyzer (E4407B, Keysight) at 4 Hz to show its evolution.

The comb power is monitored by a filter centered at 1554.5 nm with 3-dB bandwidth of 0.5 nm.

The fine tuning of the fiber cavity length at micrometer level is realized by a piezoelectric stack, whose voltage-displacement curve is calibrated by an MZI.

We choose the 2-FSR perfect soliton crystal to characterize the turnkey soliton performance in Figs. 3 and 4 due to three reasons: (i) It is easy to tell the soliton energy source comes from the SBL instead of the pump laser according to the optical spectrum analyzer with a resolution of 0.02 nm. However, it is difficult to tell the DKS's pump is from the pump or the SBL at the single-soliton state. (ii) The 2-FSR perfect soliton crystal state can convincingly demonstrate the soliton generation with deterministic state for each turnkey operation when the system parameters are correctly set; (iii) The 2-FSR perfect soliton crystal has a repetition rate of 20.18 GHz, which can still be measured and processed by our RF electronics.

**Measurement of laser phase noise and fundamental linewidth.** A self-heterodyne frequency discriminator using a fiber based unbalanced MZI and a balanced photodetector (BPD) is employed to measure the laser phase noise and fundamental linewidth. One arm of the unbalanced MZI is made of 250-m-long single mode fiber, while the other arm consists of an acousto-optic frequency shifter with frequency shift of 200 MHz and a polarization controller for high-voltage output. The FSR of the unbalanced MZI is 0.85 MHz. The two 50:50 outputs of the unbalanced MZI are connected to a BPD (PDB570C, Thorlabs) with a bandwidth of 400 MHz to reduce the impact of detector intensity fluctuations. The balanced output is then analyzed by a phase noise analyzer (NTS-1000A, RDL). The minimum fundamental linewidth that can be measured by this frequency discriminator is below 10 mHz. Other details of the laser frequency noise measurement can be found in Ref. [9].

**Measurement of comb repetition rate phase noise and timing jitter.** The RF beat note of the soliton repetition rate is directly detected by a fast photodetector (EOT, ET-3500F) and measured by an electrical spectrum analyzer (E4407B, Keysight) at 5 Hz to show its evolution. Since the soliton phase noise measurement based on fast photodetectors is limited not only by the shot noise but also the available electronics operating at high frequency, we introduce the ARMI setup [8,9,40,41] to precisely measure the phase noise of SBL soliton. Two spectral regions of 1548.5 nm ± 0.25 nm (3 dB) and 1556.5 nm ± 0.25 nm (3dB) are filtered out and sent to the interferometer. The locking bandwidth of the ARMI setup is set to be 100 Hz. Therefore, the phase noise spectrum outside the locking bandwidth above 100 Hz is measured as shown in Fig. 4b. Other details can be found in Ref. [9].

We electrically divide the RF beat note of the SBS frequency shift of 10.347 GHz by 8 times to 1.29 GHz and then measure its phase noise with a downconverter (DCR–2500A, RDL) and a phase noise analyzer (NTS-1000A, RDL).

### Data availability
All data generated or analyzed during this study are available within the paper and its Supplementary Information. Further source data will be made available on reasonable request.

### Code availability
The analysis codes will be available on reasonable request.

### Acknowledgments
We thank Professor S. A. Diddams from NIST, Boulder for fruitful discussions. We thank Dr. Bowen Li for fruitful discussion on the manuscript revision. M.N., J.B., and S.W.H. acknowledge the support from National Science Foundation (ECCS 2048202) and National Institute of Biomedical Imaging and Bioengineering (REB029541A). K.J., S. Z. and Z.X. acknowledge the support by the National Key R&D Program of China (2019YFA0705000, 2017YFA0303700), Key R&D Program of Guangdong Province (2018B030329001), Guangdong Major Project of Basic and Applied Basic Research (2020B0301030009), Zhangjiang Laboratory (ZJSP21A001), Leading-edge technology Program of Jiangsu Natural Science Foundation (BK20192001) and National Natural Science Foundation of China (51890861, 11690031, 11621091, 11627810, 11674169, 91950206).

### Author contributions
M.N. and S.W.H. conceived the idea of the experiment. M.N., J.B., and S.W.H. designed and performed the experiment, while K.J., S.Z., and Z.X. designed and fabricated the devices. M.N. and S.W.H. conducted the data analysis and wrote the manuscript. All authors contributed to the discussion and revision of the manuscript.

### Competing interests
The authors declare no competing interests.

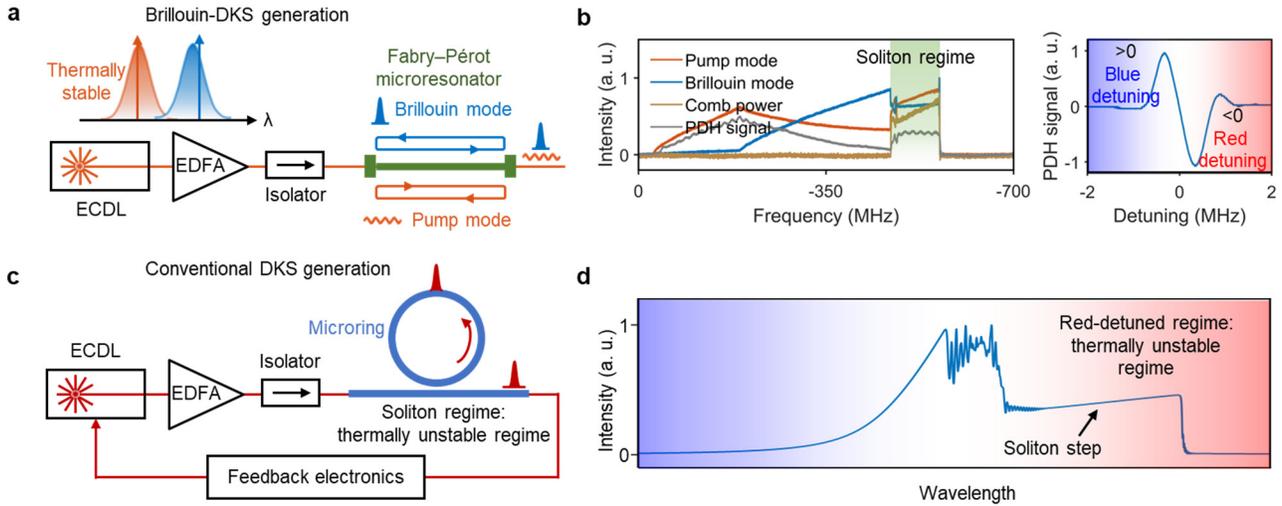

**Extended Data Fig. 1 | Different types of DKS generation and sweeping dynamics when pumping the microresonator with a tunable external cavity diode laser (ECDL, single-frequency laser).** (a)(b) Brillouin-DKS generation; (c)(d) Conventional DKS generation. (a) Schematic of Brillouin-DKS generation utilizing two-step pumping scheme. (b) Left: experimental pump PDH signal and power evolution of pump mode, Brillouin mode and combs as the ECDL frequency sweeps across the pump resonance from blue to red side. Right: frequency-calibrated pump PDH signal at low pump power. The modulation frequency of the phase modulator is 1 MHz, close to the linewidth of the high-Q microresonator. PDH signal>0: blue detuning, PDH signal<0: red detuning, PDH signal=0, zero detuning (resonance peak). During the sweeping process, stimulated Brillouin laser and Brillouin-DKS are excited. In the two-step pumping scheme, the red-detuned and comb-generating SBL serves as the direct soliton energy source, while the blue-detuned pump compensates the thermal nonlinearity and stabilizes the Brillouin-DKS without sophisticated feedback electronics. When the soliton is generated, the transmitted pump laser power increases. (c) Schematic of conventional DKS generation. (d) Simulated transmission spectrum when the pump laser is scanned from blue side to red side at a proper scanning speed depending on the thermal response. Solitons can form in the red-detuned regime seeded from modulation instability. However, complex electronics are required to stabilize the soliton since pump laser is thermally unstable in the red-detuned regime and will be kicked out immediately from the microresonator due to the perturbation [54].

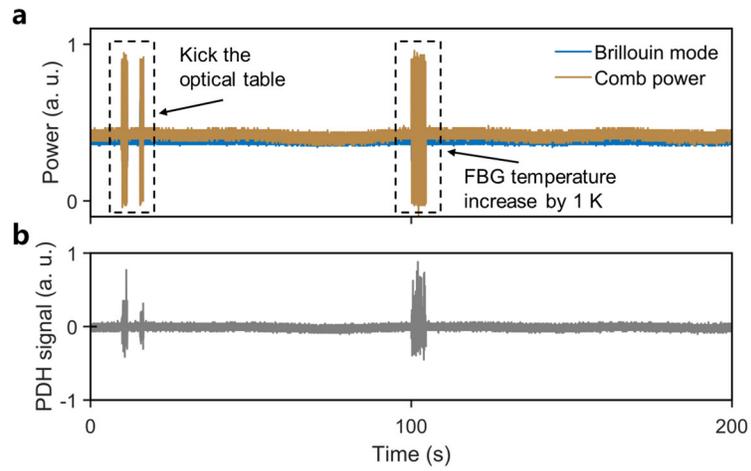

**Extended Data Fig. 2 | Soliton self-healing from perturbation**. (a) Temporal evolution of comb power and Brillouin mode power. (b) Temporal evolution of PDH signal.

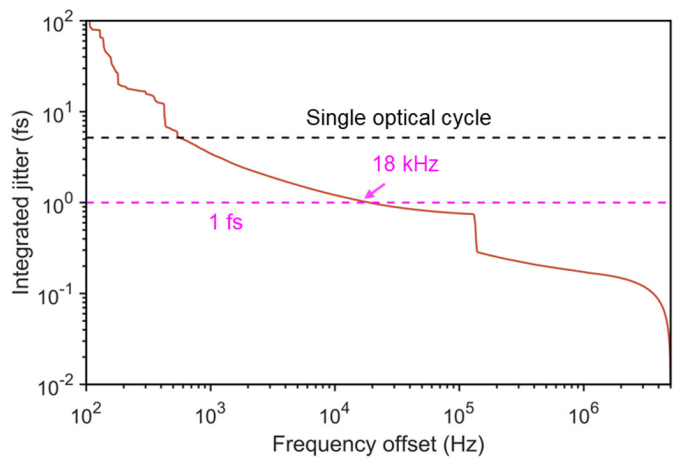

**Extended Data Fig. 3 | Integrated timing jitter.** The timing jitter integrated from 18 kHz to 1 MHz is 1 fs and timing jitter integrated from 550 Hz to 1 MHz is single optical cycle (~5.2 fs).

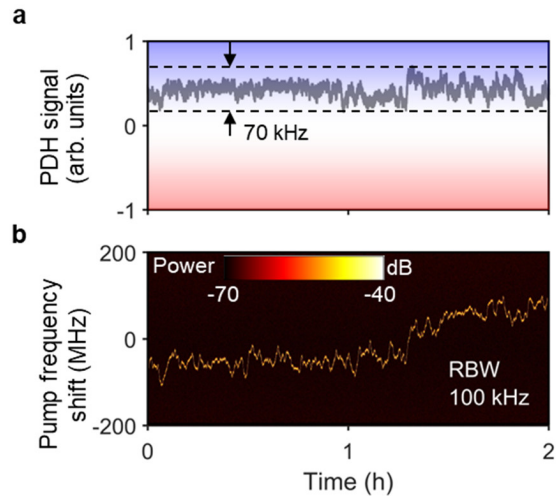

**Extended Data Fig. 4 | Long-term stability of pump detuning and pump frequency shift.** (a) Temporal evolution of pump PDH signal. (b) Temporal evolution of pump laser frequency.